\documentclass[superscriptaddress,aps,letterpaper,10pt,twocolumn,floats,showpacs,amsmath,amsfonts,amssymb,prl]{revtex4-1}

\usepackage{graphicx}
\usepackage[colorlinks=true,citecolor=blue]{hyperref}
\usepackage[caption=false]{subfig}
\usepackage{pstricks}
\usepackage{pst-node}
\usepackage{amsmath}
\usepackage{amsbsy}
\usepackage{verbatim}
\begin{document}


\title{Entropy Production in Open Systems: The Predominant Role of Intra-Environment Correlations}

\author{Krzysztof Ptaszy\'{n}ski}
\affiliation{Institute of Molecular Physics, Polish Academy of Sciences, Mariana Smoluchowskiego 17, 60-179 Pozna\'{n}, Poland}
\email{krzysztof.ptaszynski@ifmpan.poznan.pl}
\author{Massimiliano Esposito}
\affiliation{Complex Systems and Statistical Mechanics, Physics and Materials Science Research Unit, University of Luxembourg, L-1511 Luxembourg, Luxembourg}

\date{\today}

\begin{abstract}
We show that the entropy production in small open systems coupled to environments made of extended baths is predominantly caused by the displacement of the environment from equilibrium rather than, as often assumed, the mutual information between the system and the environment. The latter contribution is strongly bounded from above by the Araki-Lieb inequality, and therefore is not time-extensive, in contrast to the entropy production itself. We confirm our results with exact numerical calculations of the system-environment dynamics.
\end{abstract}

\maketitle

The emergence of thermodynamic irreversibility from the reversible dynamics is one of the most important issues of thermodynamics and statistical physics. In the context of quantum (classical) systems the problem arises from the fact that the natural candidate for the definition of the thermodynamic entropy, namely, the von Neumann (Shannon) entropy, is invariant under the unitary dynamics. In Ref.~\cite{esposito2010} this problem has been addressed by considering the joint unitary evolution of the system and the environment (which may consist of one or several baths) starting from the initially uncorrelated state ${\rho_{SE}(0)=\rho_S(0) \otimes \rho_E^\text{eq}}$; the density matrices $\rho_S$ and $\rho_{SE}$ represent here the state of the system and the joint state of the system and the environment, respectively, whereas $\rho_E^\text{eq}$ represents the Gibbs state of the environment. It was shown that the entropy production can be expressed as ${\sigma \equiv D[\rho_{SE}(t)|\rho_S(t) \rho_E^\text{eq}]}$, where ${D(\rho||\sigma)=\text{Tr}[\rho(\ln \rho-\ln \sigma)]}$ is the relative entropy (here and from here on we take ${k_B=\hbar=1}$). The second law of thermodynamics ${\sigma \geq 0}$ results then from non-negativity of the relative entropy.

A closer look shows that the entropy production can be further decomposed into two terms~\cite{esposito2010, reeb2014, uzdin2018}:
\begin{align}
\sigma=I_{SE}+D[\rho_E(t)||\rho_E^\text{eq}],
\end{align} 
where ${I_{SE}=S_S+S_E-S_{SE}}$ is the mutual information between the system and the environment and $D[\rho_E(t)||\rho_E^\text{eq}]$ is the relative entropy between the original and the final state of the environment; here ${S_i=-\text{Tr} (\rho_i \ln \rho_i)}$, with ${i \in \{S,E,SE\}}$, is the von Neumann entropy. The first term describes the system-environment correlation, whereas the second one corresponds to the displacement of the environment from equilibrium.

The natural question arising is how these terms contribute to the entropy production. It was often held~\cite{li2017, strasberg2017, chen2017, engelhardt2018, manzano2018, you2018, li2019, santos2019, bera2019} that the relative entropy $D[\rho_E(t)||\rho_E^\text{eq}]$ is negligible for large thermal reservoirs. Based on this assumption, some recent papers even directly identified the entropy production with the mutual information between the system and the environment~\cite{li2017, li2019}. In this Letter we show, however, that in small open systems driven out of equilibrium the opposite is the case. This is because the system-environment mutual information is strongly bounded from above by the inequality~\cite{jaeger2007}
\begin{align} \label{araki}
{I_{SE} \leq 2 \text{min} \{ S_S,S_E\}},
\end{align}
which is a corollary of the Araki-Lieb inequality (Theorem 2.~in Ref.~\cite{araki1970}). The maximum entropy of the system is equal to $\ln N$, where $N$ is the dimension of the Hilbert space of the system, which implies ${I_{SE} \leq 2 \ln N}$. This bound is particularly strong in systems consisting of few discrete energy levels, which are often studied in the context of quantum and stochastic thermodynamics~\cite{klages2013, seifert2012, kosloff2014, benenti2017}. The mutual information $I_{SE}$, therefore, is not a time-extensive quantity but rather saturates after a certain time, as already demonstrated numerically in Ref.~\cite{sharma2015}. In contrast, the entropy production is time-extensive in systems with a continuous current flow between the baths or systems driven by some external force. We conclude, therefore, that in such a case the entropy production is related mainly to the relative entropy contribution $D[\rho_E(t)||\rho_E^\text{eq}]$. This observation is further demonstrated by exact numerical calculations of the system-environment dynamics. We also provide a physical interpretation of the relative entropy contribution by showing that for environments made of large baths it may be attributed to generation of the mutual information between initially uncorrelated degrees of freedom in the environment.

\textit{General considerations}. --- To support our claims, let us first briefly rederive the main results of Refs.~\cite{esposito2010, reeb2014, uzdin2018}. We consider the open quantum system described by the Hamiltonian
\begin{align}
\hat{H}_{SE}(t)=\hat{H}_S(t)+\hat{H}_E+\hat{V}(t),
\end{align}
where $\hat{H}_S(t)$, $\hat{H}_E$, $\hat{V}(t)$ are the Hamiltonians of the system, environment and the interaction between the system and the environment, respectively. The Hamiltonian of the environment is assumed to be time-independent. For environments made of several baths the Hamiltonian $\hat{H}_E$ can be further decomposed as $\hat{H}_E=\sum_\alpha \hat{H}_\alpha$, where $\hat{H}_\alpha$ is the Hamiltonian of the bath $\alpha$.

Let us now consider the unitary evolution of the joint system starting from the initially uncorrelated state
\begin{align}
\rho_{SE}(0)=\rho_S(0) \otimes \rho_E^\text{eq}=\rho_S(0) \otimes \prod_{\alpha} \rho_\alpha^\text{eq}.
\end{align}
Here
\begin{align}
\rho_\alpha^\text{eq} = Z_\alpha^{-1} e^{-\beta_\alpha \left(\hat{H}_\alpha-\mu_\alpha \hat{N}_\alpha \right)}
\end{align}
is the grand canonical Gibbs state of the bath $\alpha$, where $\beta_\alpha$ and $\mu_\alpha$ are the inverse temperature and the chemical potential of the bath, respectively, $\hat{N}_\alpha$ is the particle number operator and $ {Z_\alpha=\text{Tr}\{\exp[-\beta_\alpha (\hat{H}_\alpha-\mu_\alpha \hat{N}_\alpha )]\}}$ is the partition function. Since the unitary dynamics does not change the von Neumann entropy of the joint system, i.e., ${S_{SE}(t)=S_{SE}(0)}$, the mutual information between the system and the environment in the moment $t$ can be expressed as
\begin{align} \label{mutinf}
I_{SE}=\Delta S_S+\Delta S_E \geq 0,
\end{align}
where $\Delta S=S(t)-S(0)$. The entropy change of the environment can be further decomposed as
\begin{align} \label{decomp}
\Delta S_E &= -\text{Tr} \left[ \rho_E(t) \ln \rho_E(t) \right]+\text{Tr} \left( \rho_E^\text{eq} \ln \rho_E^\text{eq} \right) \\ \nonumber &= -\sum_{\alpha} \beta_\alpha Q_\alpha-D[\rho_E(t)||\rho_E^\text{eq}].
\end{align}
Here the term
\begin{align} \label{heatcontr}
-\sum_{\alpha} \beta_\alpha Q_\alpha \equiv -\text{Tr} \left[ \rho_E(t) \ln \rho_E^\text{eq} \right]+\text{Tr} \left( \rho_E^\text{eq} \ln \rho_E^\text{eq} \right)
\end{align}
is the heat-related contribution to the change of $\Delta S_E$, with
\begin{align} \label{heat}
Q_\alpha \equiv -\text{Tr} \left\{\left[\rho_\alpha(t)-\rho_\alpha^\text{eq}\right] \left(\hat{H}_\alpha-\mu_\alpha \hat{N}_\alpha \right) \right\}
\end{align}
being the heat delivered to the system from the reservoir $\alpha$ within the time interval $[0,t]$. The second term,
\begin{align} \label{relentr}
D[\rho_E(t)||\rho_E^\text{eq}] \equiv \text{Tr} \left[\rho_E(t) \ln \rho_E(t) \right]-\text{Tr} \left[ \rho_E(t) \ln \rho_E^\text{eq} \right],
\end{align}
is the aforementioned relative entropy between the original and the final state of the environment. 

Inserting Eq.~\eqref{decomp} into Eq.~\eqref{mutinf} and rearranging terms one obtains the second law of thermodynamics
\begin{align}
\sigma \equiv \Delta S_S - \sum_{\alpha} \beta_\alpha Q_\alpha =I_{SE}+D[\rho_E(t)||\rho_E^\text{eq}] \geq 0.
\end{align}
This equation relates the standard thermodynamic definition of the entropy production to the information-theoretical quantities $I_{SE}$ and $D[\rho_E(t)||\rho_E^\text{eq}]$. As discussed before, when the entropy production significantly exceeds $2 \ln N$ it has to be related mainly to the relative entropy contribution: $\sigma \approx \sum_\alpha -\beta_\alpha Q_\alpha \approx D[\rho_E(t)||\rho_E^\text{eq}]$. This conclusion may be surprising because it was often held~\cite{li2017, strasberg2017, chen2017, engelhardt2018, manzano2018, you2018, li2019, santos2019, bera2019} that the term $D[\rho_E(t)||\rho_E^\text{eq}]$ is of the second order to the change of the density matrix of the environment ${\Delta \rho_E=\rho_E(t)-\rho_E^\text{eq}}$, and therefore can be neglected for large thermal reservoirs. However, whereas such order-of-magnitude arguments are valid for numbers, they should be applied with care when considering complex, multi-element structures, such as density matrices; this is because a sum of many small contributions can still be significant. As a matter of fact, a non-negligible value of the relative entropy contribution for an extended bath has been already numerically demonstrated in Refs.~\cite{pucci2013, goyal2019}, however, without noting the generality of this result; see also a similar observation of a non-vanishing contribution to $\Delta S_E$ not related to heat in Ref.~\cite{aurell2015}.

Furthermore, the Araki-Lieb inequality can be easily rewritten as ${\Delta S_E+S_S(t)-S_S(0) \leq 2 S_S(t)}$; thus ${\Delta S_E \leq S_S(0)+S_S(t) \leq 2 \ln N}$. This implies that the change of the von Neumann entropy of the environment is also strongly bounded from above and possibly much smaller than the heat related contribution $-\sum_\alpha \beta_\alpha Q_\alpha$, which we later demonstrate numerically. Therefore, the change of the von Neumann entropy of the isothermal environment cannot be identified with the heat taken from the environment by the relation ${\Delta S_E=-Q/T}$, as done in equilibrium thermodynamics. Instead, as follows from Eq.~\eqref{decomp}, the identity ${-Q/T=\Delta S_E+D[\rho_E(t)||\rho_E^\text{eq}]}$ holds, which clearly shows that the reservoir has been pushed away from equilibrium.

\textit{Relative entropy and inter-environment correlations}. --- This raises the question of the physical meaning of the relative entropy. Here we show that it can be, at least partially, attributed to generation of the correlation between initially uncorrelated degrees of freedom in the environment. For simplicity, let us focus on environments made of noninteracting baths described by Hamiltonians of the form
\begin{align}
\hat{H}_{\alpha}=\sum_k \epsilon_{\alpha k} c^\dagger_{\alpha k} c_{\alpha k},
\end{align}
where $c_{\alpha k}^\dagger$ ($c_{\alpha k}$) is the creation (annihilation) of the particle (boson or fermion) with the energy $\epsilon_{\alpha k}$. The thermal state of the environment can be then written as
\begin{align}
\rho_E^\text{eq} =\prod_{\alpha k} \rho_{\alpha k}^\text{eq}=\prod_{\alpha k} Z_{\alpha k}^{-1} e^{-\beta_\alpha (\epsilon_{\alpha k}-\mu_\alpha) c^\dagger_{\alpha k} c_{\alpha k}},
\end{align}
where ${\rho_{\alpha k}^\text{eq}=Z_{\alpha k}^{-1} \exp[-\beta_\alpha (\epsilon_{\alpha k}-\mu_\alpha) c^\dagger_{\alpha k} c_{\alpha k}]}$ is the equilibrium density matrix of a single level with ${Z_{\alpha k}=\text{Tr} \{\exp[-\beta_\alpha (\epsilon_{\alpha k}-\mu_\alpha) c^\dagger_{\alpha k} c_{\alpha k}]\}}$. The relative entropy of the environment can be further decomposed into two non-negative contributions as
\begin{align} \label{reldec}
D[\rho_E(t)||\rho_E^\text{eq}]=D_\text{env}+I_\text{env},
\end{align}
where
\begin{align}
D_\text{env}=\sum_{\alpha k} D[\rho_{\alpha k}(t)||\rho_{\alpha k}^\text{eq}]
\end{align}
is the sum of relative entropies of the levels and
\begin{align}
I_\text{env}=\sum_{\alpha k} S_{\alpha k}-S_E
\end{align}
is the mutual information describing the intra-environment correlations, with ${S_{\alpha k}=-\text{Tr} (\rho_{\alpha k} \ln \rho_{\alpha k})}$ being the von Neumann entropy of the level $\alpha k$ (for derivation of Eq.~\eqref{reldec}, see the Supplemental Material~\cite{supp}). We later show numerically that the second contribution may become dominant for large baths. Furthermore, in the Supplemental Material~\cite{supp} we demonstrate that, at least for noninteracting systems, the relative entropy of a single level $D[\rho_{\alpha k}(t)||\rho_{\alpha k}^\text{eq}]$ is of second order in the change of the level occupancy. Thus, the contribution $D_\text{env}$ should vanish in the thermodynamic limit in which the population of each level is only weakly perturbed. However, since the order-of-magnitude analysis can be sometimes misleading (as shown before), this latter statement should be taken with care.

\textit{Example: noninteracting resonant level}. --- We demonstrate our results on the example of a single fermionic level (denoted as $d$) coupled to two fermonic baths $\alpha \in \{L, R\}$, each containing $K$ discrete energy levels $\alpha k$ with $k \in \{1,...,K\}$. Generation of the mutual information in such a system has been already analyzed in Ref.~\cite{sharma2015}, however, without connection to thermodynamics. The Hamiltonian of the system reads
\begin{align} \label{hamilt}
\hat{H}&=\epsilon_d c_d^\dagger c_d+\sum_{\alpha k} \epsilon_{\alpha k} c^\dagger_{\alpha k} c_{\alpha k} +\sum_{\alpha k} \left(t_{\alpha k} c_d^\dagger c_{\alpha k} + \text{h.c.} \right),
\end{align}
where ${i,j \in \{ d, L1,...,LK,R1,...,RK\}}$ denote the sites, $c_i^\dagger$ ($c_i$) is the creation (annihilation) operator of the particle on the site $i$ and  $t_{\alpha k}$ is the tunnel coupling between the central level and the site $\alpha k$. We further take the energy levels of the baths to be equally spaced, with ${\epsilon_{\alpha,k+1}=\epsilon_{\alpha k}+\Delta \epsilon}$, where ${\Delta \epsilon=W/(K-1)}$, with $W$ being the bandwidth of the reservoirs. We also express the tunneling elements as ${t_{\alpha k}=\sqrt{\Gamma_\alpha \Delta \epsilon/(2 \pi)}}$, where $\Gamma_\alpha$ is the coupling strength to the bath $\alpha$. 

The exact description of many-body systems is usually not possible since the rank of the density matrix, and thus the computational complexity, increases exponentially with the size of the system. However, since the Hamiltonian~\eqref{hamilt} is quadratic, the state of the system can be fully described by the two-point correlation matrix $\mathcal{C}$ with the matrix elements defined as~\cite{peschel2003}
\begin{align}
\mathcal{C}_{ij}(t) = \text{Tr} \left[ c_i^\dagger c_j \rho_{SE}(t)\right].
\end{align}
The rank of the correlation matrix $2K+1$ increases only linearly with the size of the system, which makes the exact description of the system numerically tractable. 

The evolution of the correlation matrix is described by the equation~\cite{eisler2012}
\begin{align} \label{evolv}
\mathcal{C}(t) =e^{i \mathcal{H} t} \mathcal{C}(0) e^{-i \mathcal{H} t},
\end{align}
where $\mathcal{C}(0)$ is the initial state and $\mathcal{H}$ is the matrix containing the Hamiltonian elements $\mathcal{H}_{ij}$ with ${\mathcal{H}_{ii}=\epsilon_i}$ and ${\mathcal{H}_{\alpha k,d}=\mathcal{H}_{d,\alpha k}^*=t_{\alpha k}}$ (we rederive this equation in the Supplemental Material~\cite{supp}). The initial correlation matrix can be expressed as
\begin{align}
\mathcal{C}(0)=\text{diag} \left[n_d(0),  n_{L1},...,n_{LK},n_{R1},...,n_{RK} \right],
\end{align}
where $n_d(0)$ is the initial occupancy of the central level and ${n_{\alpha k}=f[\beta_\alpha (\epsilon_{\alpha k}-\mu_\alpha)]}$ are the thermal occupancies of the sites $\alpha k$, with $f(x)$ being the Fermi distribution.

The von Neumann entropy of the subsystem $\mathcal{G}$ can be calculated as~\cite{sharma2015}
\begin{align}
S_\mathcal{G} =-\sum_{\sigma} \left[ C_\sigma \ln C_\sigma + \left(1-C_\sigma \right) \ln \left(1-C_\sigma \right) \right],
\end{align}
where $C_\sigma$ are the eigenvalues of the reduced correlation matrix $\mathcal{C}_\mathcal{G}$ defined within the subspace $\mathcal{G}$; for example, $\mathcal{C}_E$ is the submatrix of the correlation matrix containing all the elements $\mathcal{C}_{ij}$ with $i,j \neq d$. In particular, the von Neumann entropy of a single level $i$ equals just ${S_{i} =- \mathcal{C}_{ii} \ln \mathcal{C}_{ii} - (1-\mathcal{C}_{ii} ) \ln (1-\mathcal{C}_{ii} )}$ and $S_S=S_d$. The heat taken from the bath $\alpha$ is expressed as
\begin{align}
Q_\alpha=-\sum_{k=1}^K \left[ \mathcal{C}_{k \alpha,k\alpha}(t)-\mathcal{C}_{k \alpha,k\alpha}(0) \right] \left(\epsilon_{k \alpha}-\mu_\alpha \right),
\end{align}
which is equivalent to Eq.~\eqref{heat}. Using Eq.~\eqref{decomp} one may further calculate the relative entropy of the environment as $D[\rho_E(t)||\rho_E^\text{eq}]=-\beta_L Q_L-\beta_R Q_R-\Delta S_E$.

%
\begin{figure}
	\centering
	\includegraphics[width=0.90\linewidth]{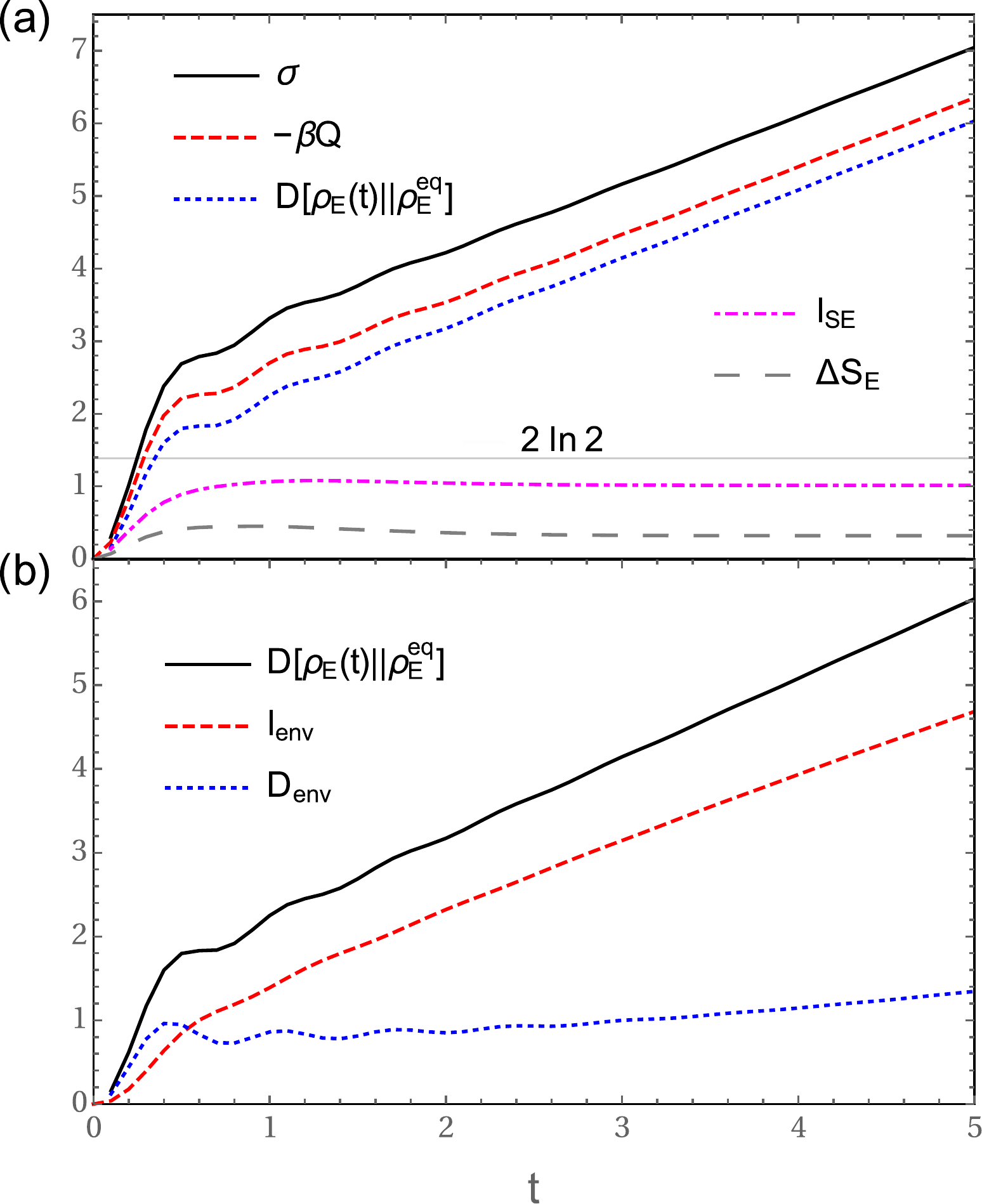}		
	\caption{(a) The entropy production $\sigma$, the heat-related contribution to the entropy production ${-\beta Q=-\beta (Q_L+Q_R)}$, the relative entropy of the environment $D[\rho_E(t)||\rho_E^\text{eq}]$, the system-environment mutual information $I_{SE}$ and the change of the von Neumann entropy of the environment $\Delta S_E$ as a function of time for ${n_d(0)=0}$, ${\Gamma_L=\Gamma_R=1/2}$, ${\mu_L=-\mu_R=1}$, ${\beta_L=\beta_R=\beta=3}$, ${\epsilon_{\alpha k}=(k-1)\Delta \epsilon-W/2}$, ${\Delta \epsilon=W/(K-1)}$, ${W=20}$ and ${K=256}$. (b) Contributions to the relative entropy $D[\rho_E(t)||\rho_E^\text{eq}]$ for parameters as in (a).}
	\label{fig:entr}
\end{figure}
%

Let us now analyze the entropy production resulting from the current flow induced by the difference of chemical potentials (voltage) ${V=\mu_L-\mu_R}$. In Fig.~\ref{fig:entr}(a) we present the time evolution of the analyzed quantities for a given number of sites ${K=256}$. One may observe that the mutual information is saturated after a time ${t \approx 1}$ and does not exceed $2 \ln 2$, in agreement with the Araki-Lieb inequality. As shown in Ref.~\cite{sharma2015}, the bound ${I_{SE} \leq 2 \ln 2}$ becomes tight for high voltages $V$. Furthermore, the change of the von Neumann entropy of the environment $\Delta S_E$ saturates in a similar way. In contrast, the entropy production $\sigma$ significantly exceeds $2 \ln 2$ and consists mostly of the relative entropy contribution $D[\rho_E(t)||\rho_E^\text{eq}]$. For $t \approx 2$ the system reaches the asymptotic long-time state in which the entropy production, heat and the relative entropy increase monotonously. This long-time state is approximately equivalent to the steady state calculated in the thermodynamic limit; due to the finite size of the baths the entropy production is, however, finally saturated for ${t \approx 80}$ (see the Supplemental Material~\cite{supp}). 

In Fig.~\ref{fig:entr}(b) we display different contributions to the relative entropy $D[\rho_E(t)||\rho_E^\text{eq}]$. For a given size of the bath the dominant contribution to the relative entropy, and thus the entropy production, is the mutual information between degrees of freedom of the environment $I_\text{env}$; the term $D_\text{env}$ is, however, also non-negligible and time-extensive. As shown in the Supplemental Material~\cite{supp}, the contribution $I_\text{env}$ is related both to the correlation between the baths and the intra-bath correlations.

%
\begin{figure}
	\centering
	\includegraphics[width=0.90\linewidth]{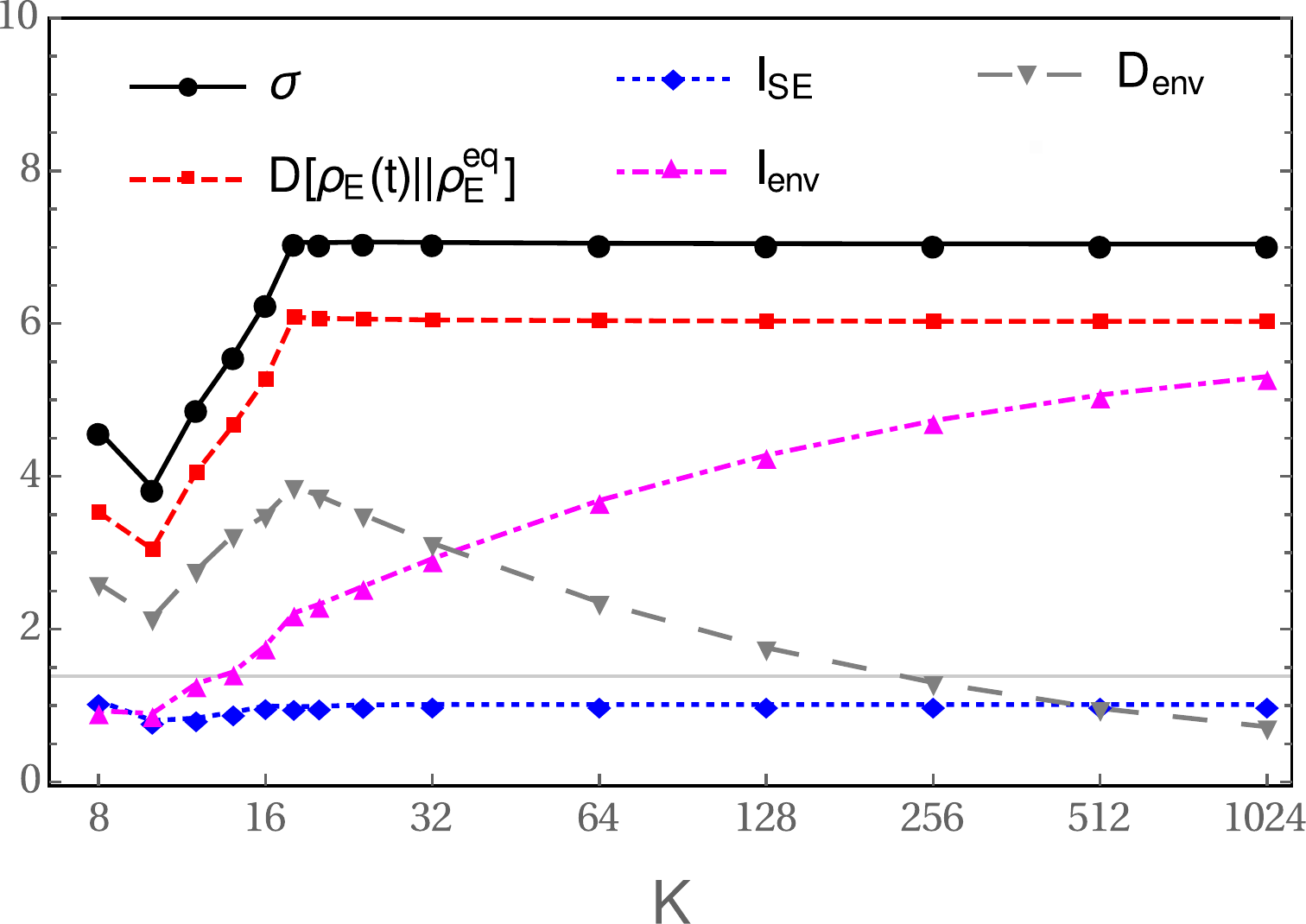}		
	\caption{The thermodynamic quantities as a function of the number of sites $K$ for ${t=5}$ and other parameters as in Fig.~\ref{fig:entr}. Results denoted by points, lines shown for eye guidance.}
	\label{fig:scaling}
\end{figure}
%

In Fig.~\ref{fig:scaling} we present the dependence of the analyzed quantities on the size of the bath for a fixed time ${t=5}$. One can observe a sharp transition at ${K \approx 18}$ which results from a cross-over in the dynamics of the system: for ${K \gtrapprox 18}$ the entropy production grows monotonously at ${t=5}$, whereas for ${K \lessapprox 18}$ it has already saturated due to the finite size of the bath (see the Supplemental Material~\cite{supp} for details). In particular, for ${K \gtrapprox 18}$ the entropy production, the relative entropy of the environment and the system-environment mutual information become independent of the number of sites. This shows that the importance of the contribution $D[\rho_E(t)||\rho_E^\text{eq}]$ to the entropy production is not related to the size of the bath. However, the contributions of the terms $D_\text{env}$ and $I_\text{env}$ to the relative entropy of the environment change with the size of the bath. For ${K \gtrapprox 18}$ the term $D_\text{env}$ decreases with the number of sites; this may be described by a power law ${D_\text{env} \propto K^{-x}}$ with ${x \approx 0.38}$. One may expect, therefore, that in the thermodynamic limit the term $D_\text{env}$ vanishes and the mutual information of the intra-environment correlations $I_\text{env}$ becomes the predominant contribution to the entropy production. This can be explained in the following way: The term $D_\text{env}$ is related to the deviation of the level occupancies (diagonal elements of the correlation matrix $\mathcal{C}$) from the equilibrium ones, which becomes negligible when the baths become large (i.e., the occupancies stay thermalized during the system evolution). On the other hand, the mutual information $I_\text{env}$ is associated with the creation of two-point correlations $\langle c_i^\dagger c_j \rangle$ (off-diagonal elements of the correlation matrix) which are not present in the thermal state; these correlations are responsible for the entropy production.

\textit{Final remarks}. --- We reemphasize that our paper focuses on the situation when the entropy production is time-extensive, such that it significantly exceeds $2 \ln N$ it the long-time limit. Whenever this assumption holds, the result relating the entropy production to the displacement of the environment from equilibrium $D[\rho_E(t)||\rho_E^\text{eq}]$ rather than the system-environment mutual information $I_{SE}$ is general and solid (since the Araki-Lieb inequality is universally valid). Therefore, although our numerical analysis focuses on the system with a time-independent Hamiltonian, this result holds also for externally driven systems, as well as setups described within the repeated interaction framework~\cite{strasberg2017, scarani2002, karevski2009, barra2015, chiara2018, cusumano2018} (in which the environment is made of independently prepared units interacting sequentially with the system).  When, on the other hand, the entropy production is saturated at value smaller or comparable to $2 \ln N$ (which may be true, e.g., for systems undergoing thermalization~\cite{cusumano2018} or short interaction quench~\cite{goyal2019}) the relative importance of the terms $D[\rho_E(t)||\rho_E^\text{eq}]$ and $I_{SE}$ may be not given by any general rule; instead, it may depend on details of the system-environment dynamics. This will be the topic of future study.

It is also not entirely clear whether in the thermodynamic limit the relative entropy of the environment can be always identified with the mutual information between degrees of freedom in the environment. Whereas it appears to be true for noninteracting baths composed of many levels continuously coupled to the system, the situation can be different for interacting environments or setups described within the repeated interaction framework. Therefore, our work may motivate further studies to better understand the mechanisms controlling the different contributions to the displacement of the environment from equilibrium. These issues, beside their fundamental importance, may also have implications for engineering environments to control dissipation.

\begin{acknowledgments}
	K. P. is supported by the National Science Centre, Poland, under the project Preludium 14 (No.~2017/27/N/ST3/01604) and the doctoral scholarship Etiuda 6 (No.~2018/28/T/ST3/00154). M. E. is supported by the European Research Council project NanoThermo (ERC-2015-CoG
	Agreement No. 681456).
\end{acknowledgments}


\begin{thebibliography}{}
\bibitem{esposito2010}
M. Esposito, K. Lindenberg, and C. Van den Broeck, Entropy production as correlation between system and reservoir, \href{\doibase 10.1088/1367-2630/12/1/013013}{New J. Phys. \textbf{12}, 013013 (2010)}.

\bibitem{reeb2014}
D. Reeb and M. M. Wolf, An improved Landauer principle with finite-size corrections, \href{\doibase 10.1088/1367-2630/16/10/103011}{New J. Phys. \textbf{16}, 103011 (2014)}.

\bibitem{uzdin2018}
R. Uzdin, The Second Law and Beyond in Microscopic Quantum Setups, in
\href{\doibase 10.1007/978-3-319-99046-0_28}{ \textit{Thermodynamics in the Quantum Regime}, edited by F. Binder, L. Correa, J. Anders, and G. Adesso, Fundamental Theories of Physics, Vol. 195 (Springer, Cham, 2018), p. 681}.

\bibitem{strasberg2017}
P. Strasberg, G. Schaller, T. Brandes, and M. Esposito, Quantum and Information Thermodynamics: A Unifying Framework Based on Repeated Interactions, \href{\doibase 10.1103/PhysRevX.7.021003}{Phys. Rev. X \textbf{7}, 021003 (2017)}. 

\bibitem{chen2017}
H.-B. Chen, G.-Y. Chen, and Y.-N. Chen, Thermodynamic description of non-Markovian information flux of nonequilibrium open quantum systems, \href{\doibase 10.1103/PhysRevA.96.062114}{Phys. Rev. A \textbf{96}, 062114 (2017)}. 

\bibitem{li2017}
S.-W. Li, Production rate of the system-bath mutual information, \href{\doibase 10.1103/PhysRevE.96.012139}{Phys. Rev. E \textbf{96}, 012139 (2017)}. 

\bibitem{engelhardt2018}
G. Engelhardt and G. Schaller, Maxwell's demon in the quantum-Zeno regime and beyond, \href{\doibase 10.1088/1367-2630/aaa38d}{New J. Phys. \textbf{20}, 023011 (2018)}. 

\bibitem{manzano2018}
G. Manzano, J. M. Horowitz, and J. M. R. Parrondo, Quantum Fluctuation Theorems for Arbitrary Environments: Adiabatic and Nonadiabatic Entropy Production, \href{\doibase 10.1103/PhysRevX.8.031037}{Phys. Rev. X \textbf{8}, 031037 (2018)}. 

\bibitem{you2018}
Y.-N. You and S.-W. Li, Entropy dynamics of a dephasing model in a squeezed thermal bath, \href{\doibase 10.1103/PhysRevA.97.012114}{Phys. Rev. A \textbf{97}, 012114 (2018)}. 

\bibitem{li2019}
S.-W. Li, The correlation production in thermodynamics, \href{\doibase 10.3390/e21020111}{Entropy \textbf{21}, 111 (2019)}. 

\bibitem{santos2019}
J. P. Santos, L. C. C\'{e}leri, G. T. Landi, and M. Paternostro, The role of quantum coherence in non-equilibrium entropy production, \href{\doibase 10.1038/s41534-019-0138-y}{npj Quantum Inf. \textbf{5}, 23 (2019)}.

\bibitem{bera2019}
M. N. Bera, A. Rieral, M. Lewenstein, Z. Baghali Khanian, and A. Winter, Thermodynamics as a Consequence of Information Conservation, \href{\doibase 10.22331/q-2019-02-14-121}{Quantum \textbf{3}, 121 (2019)}.

\bibitem{jaeger2007}
G. Jaeger, \textit{Quantum Information: An Overview} (Springer, New York, 2007).

\bibitem{araki1970}
H. Araki and E. H. Lieb, Entropy inequalities, \href{\doibase 10.1007/BF01646092}{Commun. Math. Phys. \textbf{18}, 160 (1970)}.

\bibitem{klages2013}
\textit{Nonequilibrium Statistical Physics of Small Systems: Fluctuation Relations and Beyond}, ed. by R. Klages, W. Just, C. Jarzynski (Wiley-VCH, Weinheim, 2013).

\bibitem{seifert2012}
U. Seifert, Stochastic thermodynamics, fluctuation theorems and molecular machines, \href{\doibase 10.1088/0034-4885/75/12/126001}{Rep. Prog. Phys. \textbf{75}, 126001 (2012)}.

\bibitem{kosloff2014}
R. Kosloff and A. Levy, Quantum Heat Engines and Refrigerators: Continuous Devices, \href{\doibase 10.1146/annurev-physchem-040513-103724}{Annu. Rev. Phys. Chem. \textbf{65}, 365 (2014)}.

\bibitem{benenti2017}
G. Benenti, G. Casati, K. Saito, and R. S. Whitney, Fundamental aspects of steady-state conversion of heat to work at the nanoscale, \href{\doibase 10.1016/j.physrep.2017.05.008}{Phys. Rep. \textbf{694}, 1 (2017)}.

\bibitem{sharma2015}
A. Sharma and E. Rabani, Landauer current and mutual information, \href{\doibase 10.1103/PhysRevB.91.085121}{Phys. Rev. B \textbf{91}, 085121 (2015)}.

\bibitem{pucci2013}
L. Pucci, M. Esposito, and L. Peliti, Entropy production in quantum Brownian motion, \href{\doibase 10.1088/1742-5468/2013/04/P04005}{J. Stat. Mech. (2013) P04005}.

\bibitem{goyal2019}
K. Goyal, X. He, and R. Kawai, Entropy production of a small quantum system under strong coupling with an environment: A computational experiment, \href{https://arxiv.org/abs/1904.07508}{arXiv:1904.07508}.

\bibitem{aurell2015}
E. Aurell and R. Eichorn, On the von Neumann entropy of a bath linearly coupled to a driven quantum system, \href{\doibase 10.1088/1367-2630/17/6/065007}{New J. Phys. \textbf{17}, 065007 (2015)}.	

\bibitem{supp} 
See the Supplemental Material at [], which includes Refs.~\cite{strasberg2017, eisler2012, ding2009, butcher1990, benenti2017, nazarov2009, klich2009, song2011, song2012}, for the derivation of Eqs.~\eqref{reldec} and~\eqref{evolv}, and additional discussions.

\bibitem{eisler2012}
V. Eisler and I. Peschel, On entanglement evolution across defects in critical chains, \href{\doibase 10.1209/0295-5075/99/20001}{Europhys. Lett. \textbf{99}, 20001 (2012)}.

\bibitem{ding2009}
W. Ding and K. Yang, Entanglement entropy and mutual information in Bose-Einstein condensates, \href{\doibase 10.1103/PhysRevA.80.012329}{Phys. Rev. A \textbf{80}, 012329 (2009)}.

\bibitem{butcher1990}
P. N. Butcher, Thermal and electrical transport formalism for electronic microstructures with many terminals, \href{\doibase 10.1088/0953-8984/2/22/008}{J. Phys.: Condens. Matter \textbf{2}, 4869 (1990)}. 

\bibitem{nazarov2009}
Yu. V. Nazarov and Ya. M. Blanter, \textit{Quantum Transport} (Cambridge University Press, Cambridge, 2009).

\bibitem{klich2009}
I. Klich and L. Levitov, Quantum Noise as an Entanglement Meter, \href{\doibase 10.1103/PhysRevLett.102.100502}{Phys. Rev. Lett. \textbf{102}, 100502 (2009)}.

\bibitem{song2011}
H. F. Song, C. Flindt, S. Rachel, I. Klich, and K. Le Hur, Entanglement entropy from charge statistics: Exact relations for noninteracting many-body systems, \href{\doibase 10.1103/PhysRevB.83.161408}{Phys. Rev. B \textbf{83}, 161408(R) (2011)}.

\bibitem{song2012}
H. F. Song, S. Rachel, C. Flindt, I. Klich, and K. Le Hur, Bipartite fluctuations as a probe of many-body entanglement, \href{\doibase 10.1103/PhysRevB.85.035409}{Phys. Rev. B \textbf{85}, 035409 (2012)}.

\bibitem{peschel2003}
I. Peschel, Calculation of reduced density matrices from correlation functions, \href{\doibase 10.1088/0305-4470/36/14/101}{J. Phys. A: Math. Gen. \textbf{36}, L205 (2003)}.

\bibitem{scarani2002}
V. Scarani, M. Ziman, P. \u{S}telmachovi\u{c}, N. Gisin, and V. Bu\u{z}ek, Thermalizing Quantum Machines: Dissipation and Entanglement, \href{\doibase 10.1103/PhysRevLett.88.097905}{Phys. Rev. Lett. \textbf{88}, 097905 (2002)}.

\bibitem{karevski2009}
D. Karevski and T. Platini, Quantum Nonequilibrium Steady States Induced by Repeated Interactions, \href{\doibase 10.1103/PhysRevLett.102.207207}{Phys. Rev. Lett. \textbf{102}, 207207 (2009)}.

\bibitem{barra2015}
F. Barra, The thermodynamic cost of driving quantum systems by their boundaries, \href{\doibase 10.1038/srep14873}{Sci. Rep. \textbf{5}, 14873 (2015)}.

\bibitem{chiara2018}
G. De Chiara, G. Landi, A. Hewgill, B. Reid, A. Ferraro, A. J. Roncaglia, and M. Antezza, Reconciliation of quantum local master equations with thermodynamics, \href{\doibase 10.1088/1367-2630/aaecee}{New J. Phys. \textbf{20}, 113024 (2018)}.

\bibitem{cusumano2018}
S. Cusumano, V. Cavina, M. Keck, A. De Pasquale, and V. Giovannetti, Entropy production and asymptotic factorization via thermalization: a collisional model approach, \href{\doibase 10.1103/PhysRevA.98.032119}{Phys. Rev. A \textbf{98}, 032119 (2018)}.


\end{thebibliography}
\end{document}


	
\title{Supplemental Material to: Entropy Production in Open Systems: The Predominant Role of Intra-Environment Correlations}
	
\author{Krzysztof Ptaszy\'{n}ski}
\affiliation{Institute of Molecular Physics, Polish Academy of Sciences, Mariana Smoluchowskiego 17, 60-179 Pozna\'{n}, Poland}
\email{krzysztof.ptaszynski@ifmpan.poznan.pl}
\author{Massimiliano Esposito}
\affiliation{Physics and Materials Science Research Unit, University of Luxembourg, L-1511 Luxembourg, Luxembourg}
	
\date{\today}

\maketitle

\section*{Supplemental material}
This appendix contains in the following order:

\begin{itemize}
	\item Derivation of Eq.~(14) from the main text [Sec.~\ref{sec:derdecrel}]
	\item Order-of-magnitude analysis of the term $D_\text{env}$ [Sec.~\ref{sec:ins}]
	\item Derivation of Eq.~(19) from the main text [Sec.~\ref{sec:derdyn}]
	\item Comparison with the results obtained in the thermodynamic limit for the steady state [Sec.~\ref{sec:compst}] 
	\item Analysis of the scale dependence of the transient dynamics [Sec.~\ref{sec:sc}]
	\item Analysis of the mutual information between the baths [Sec.~\ref{sec:infdec}]
\end{itemize}

\subsection{Derivation of Eq.~(14)} \label{sec:derdecrel}
Here we derive the splitting of the relative entropy into contributions $D_\text{env}$ and $I_\text{env}$. Let us first rewrite Eq.~(10) from the main text as
\begin{align} \nonumber
& D[\rho_E(t)||\rho_E^\text{eq}] =  \text{Tr} \left[\rho_E(t) \ln \rho_E(t) \right]-\text{Tr} \left[ \rho_E(t) \ln \rho_E^\text{eq} \right] \\
&+\sum_{\alpha k} \text{Tr} \left[ \rho_{\alpha k}(t) \ln \rho_{\alpha k}(t) \right]-\sum_{\alpha k} \text{Tr} \left[ \rho_{\alpha k}(t) \ln \rho_{\alpha k}(t) \right],
\end{align}
where we just added and subtracted $\sum_{\alpha k} \text{Tr} \left[ \rho_{\alpha k}(t) \ln \rho_{\alpha k}(t) \right]$; here, as in the main text, $\rho_{\alpha k}$ is the reduced density matrix of a single level. Then, Eq.~(14) from the main text can be readily obtained by splitting the relative entropy as
\begin{align}
D[\rho_E(t)||\rho_E^\text{eq}]=D_\text{env}+I_\text{env},
\end{align}
with
\begin{align} \nonumber \label{infenv}
I_\text{env}&=-\sum_{\alpha k} \text{Tr} \left[ \rho_{\alpha k}(t) \ln \rho_{\alpha k}(t) \right]+ \text{Tr} \left[\rho_E(t) \ln \rho_E(t) \right] \\ &=\sum_{\alpha k} S_{\alpha k}-S_E,
\end{align}
and
\begin{align} \label{denv} \nonumber
D_\text{env} &=\sum_{\alpha k} \text{Tr} \left[ \rho_{\alpha k}(t) \ln \rho_{\alpha k}(t) \right]-\text{Tr} \left[ \rho_E(t) \ln \rho_E^\text{eq} \right] \\ \nonumber
&=\sum_{\alpha k} \text{Tr} \left[ \rho_{\alpha k}(t) \ln \rho_{\alpha k}(t) \right]- \sum_{\alpha k} \text{Tr} \left[ \rho_{\alpha k}(t) \ln \rho_{\alpha k}^\text{eq} \right]\\ &=\sum_{\alpha k} D[\rho_{\alpha k}(t)||\rho_{\alpha k}^\text{eq}].
\end{align}
One can verify the validity of the second step in Eq.~\eqref{denv} step by noting
\begin{align} \nonumber
&\text{Tr} \left[ \rho_E(t) \ln \rho_E^\text{eq} \right] \\ \nonumber &=-\sum_{\alpha} \beta_\alpha \text{Tr} \left[ \rho_E(t) \left(\hat{H}_\alpha-\mu_\alpha \hat{N}_\alpha \right) \right]-\ln \prod_{\alpha} Z_\alpha \\ \nonumber
&=-\beta_\alpha \sum_{\alpha k} n_{\alpha k}(t) (\epsilon_{\alpha k}-\mu_\alpha) -\ln \prod_{\alpha k} Z_{\alpha k} \\ \nonumber 
&=-\beta_\alpha \sum_{\alpha k} (\epsilon_{\alpha k}-\mu_\alpha) \text{Tr} \left[c^\dagger_{\alpha k} c_{\alpha k} \rho_{\alpha k} (t) \right]-\sum_{\alpha k} \ln Z_{\alpha k} \\
&=\sum_{\alpha k} \text{Tr} \left[ \rho_{\alpha k}(t) \ln \rho_{\alpha k}^\text{eq} \right],
\end{align}
where $n_{\alpha k}(t)$ is the occupancy of the level $\alpha k$ in the moment $t$.

\subsection{Order-of-magnitude analysis of the term $D_\text{env}$} \label{sec:ins}
Here we analyze the magnitude of the term $D_\text{env}$ for small changes of the bath level populations. Let us first notice that [in analogy to Eq.~(7) from the main text] the change of the von Neumann entropy of a single level can be written as
\begin{align} \label{entrdec}
\Delta S_{\alpha k} = -\beta_\alpha Q_{\alpha k} -D[\rho_{\alpha k}(t)||\rho_{\alpha k}^\text{eq}],
\end{align}
where ${Q_{\alpha k} = -\Delta n_{\alpha k} (\epsilon_{\alpha k}-\mu_\alpha)}$ is the heat taken from the site $\alpha k$, with ${\Delta n_{\alpha k}=n_{\alpha k}(t)-n_{\alpha k}^\text{eq}}$ being the change of the level occupancy $n_{\alpha k}(t)$ with respect to the equilibrium occupancy $n_{\alpha k}^\text{eq}$. For fermionic levels without coherences between the Fock states the von Neumann entropy of the level  $\alpha k$ clearly depends only on the level occupancy:
\begin{align} 
S_{\alpha k}=-n_{\alpha k} \ln n_{\alpha k}-\left(1-n_{\alpha k} \right) \ln \left(1-n_{\alpha k} \right).
\end{align}
The change of the von Neumann entropy can be then written as
\begin{align} \label{entrch}
\Delta S_{\alpha k}=&-\left(\Delta n_{\alpha k}+n_{\alpha k}^\text{eq} \right) \ln \left(\Delta n_{\alpha k}+n_{\alpha k}^\text{eq} \right) \\ \nonumber
&-\left(1-\Delta n_{\alpha k}-n_{\alpha k}^\text{eq} \right) \ln \left(1-\Delta n_{\alpha k}-n_{\alpha k}^\text{eq} \right) \\ \nonumber
&+ n_{\alpha k}^\text{eq} \ln n_{\alpha k}^\text{eq}+ \left(1-n_{\alpha k}^\text{eq} \right) \ln \left(1-n_{\alpha k}^\text{eq} \right) \\ \nonumber
=& \Delta n_{\alpha k} \ln \left( \frac{1-n_{\alpha k}^\text{eq}}{n_{\alpha k}^\text{eq}} \right) + \mathcal{O}(\Delta n_{\alpha k}^2) \\ \nonumber
=& -\beta_\alpha Q_{\alpha k}+ \mathcal{O}(\Delta n_{\alpha k}^2),
\end{align}
where in the second step we applied the Taylor expansion around ${\Delta n_{\alpha k}=0}$ and in the third step we inserted the Fermi distribution ${n_{\alpha k}^\text{eq}=f[\beta_\alpha (\epsilon_{\alpha k}-\mu_\alpha)]}$. Comparing Eq.~\eqref{entrdec} with Eq.~\eqref{entrch} one finds that $D[\rho_{\alpha k}(t)||\rho_{\alpha k}^\text{eq}]$ is in the second order in the change of the level population, and thus becomes negligible for small $\Delta n_{\alpha k}$. For baths consisting of many levels coupled to the system the change of the level population will be indeed small. This explains why the contribution $D_\text{env}$ decreases when the size of the bath increases (Fig.~2 in the main text).

For bosonic systems the situation is less clear. However, it was proved~\cite{ding2009} that for initially thermal noninteracting bosonic systems the von Neumann entropy of a single level is also fully given by its population and reads
\begin{align}
S_{\alpha k}=\left(1+n_{\alpha k} \right) \ln \left(1+n_{\alpha k} \right)-n_{\alpha k} \ln n_{\alpha k}.
\end{align}
Repeating the same procedure as for fermionic systems and applying ${n_{\alpha k}^\text{eq}=n_B[\beta_\alpha (\epsilon_{\alpha k}-\mu_\alpha)]}$, where $n_B(x)$ is the Bose-Einstein distribution, one arrives at the same conclusion.

One should be aware that the change of the level population $\Delta n_{\alpha k}$ does not have to be always small. For example, for systems described within the repeated interaction framework~\cite{strasberg2017} the change of state of the unit interacting with the system can be significant when the interaction between the system and the unit is strong; in such a case the contribution $D_\text{env}$ may be important.

\subsection{Derivation of Eq.~(19)} \label{sec:derdyn}
Here we rederive the equation of motion for the correlation matrix [Eq.~(19) from the main text], which was previously applied in Ref.~\cite{eisler2012}. Let us consider the evolution of the correlation matrix induced by a generic fermionic quadratic Hamiltonian of the form
\begin{align}
\hat{H}=\sum_{ij} \mathcal{H}_{ji} c_i^\dagger c_j.
\end{align}
The dynamics of a single matrix element can be described as
\begin{align} \nonumber
d_t \mathcal{C}_{ij}&=\text{Tr} \left(c_i^\dagger c_j d_t \rho_{SE} \right)=-i \text{Tr} \left(c_i^\dagger c_j \left[\hat{H}, \rho_{SE} \right] \right) \\ \nonumber
& =-i \sum_{kl} \mathcal{H}_{lk} \text{Tr} \left(c_i^\dagger c_j \left[c_k^\dagger c_l, \rho_{SE} \right] \right) \\ \nonumber
& = i \sum_{kl} \mathcal{H}_{lk} \text{Tr} \left [\left(c_k^\dagger c_l c_i^\dagger c_j-c_i^\dagger c_j c_k^\dagger c_l \right) \rho_{SE} \right ] \\ \nonumber
& =i \sum_{kl} \mathcal{H}_{lk} \text{Tr} \left [\left(\delta_{il} c_k^\dagger c_j-\delta_{jk} c_i^\dagger c_l \right) \rho_{SE} \right ] \\ 
&= i\sum_{k} \mathcal{H}_{ik} \mathcal{C}_{kj}-i\sum_l \mathcal{C}_{il} \mathcal{H}_{lj} =i \left[\mathcal{H}, \mathcal{C} \right]_{ij}.
\end{align}
Here in the fourth step we used the cyclic property of the trace, in the fifth step -- the commutation properties of the creation and annihilation operators, and in the last step we applied the definition of the matrix product. As a result, the evolution of the correlation matrix is given by the formula resembling the equation of motion for operators in the Heisenberg picture
\begin{align} \label{eqmot}
d_t \mathcal{C} = i \left[\mathcal{H}, \mathcal{C} \right].
\end{align}
Using this analogy, for a time-independent Hamiltonian Eq.~\eqref{eqmot} can be solved as
\begin{align}
\mathcal{C}(t) =e^{i \mathcal{H} t} \mathcal{C}(0) e^{-i \mathcal{H} t},
\end{align}
which is Eq.~(19) from the main text.

\subsection{Comparison with the thermodynamic limit for the steady state} \label{sec:compst}
The correlation matrix formalism considered in the main text is confined to systems with finite baths. Here we demonstrate, however, that the asymptotic long time heat currents calculated using this method converge to the steady state currents calculated in the thermodynamic limit by means of the Landauer-B\"{u}ttiker transmission formalism. Within the latter approach the heat current flowing to the lead $\alpha$ for a noninteracing system attached to two fermionic leads is expressed as~\cite{butcher1990, benenti2017}
\begin{align} \label{hatcurtr}
\dot{Q}_{\alpha,\text{TR}} = \int_{-\infty}^{\infty} \frac{d\omega}{2 \pi} \Delta_\alpha T(\omega) \left[f_L(\omega)-f_R (\omega) \right],
\end{align}
where $T(\omega)$ is the transmission function, ${f_\alpha(\omega)=f[\beta_\alpha (\omega-\mu_\alpha)]}$ is the Fermi distribution of the lead $\alpha$, ${\Delta_L=\omega-\mu_L}$ and ${\Delta_R=\mu_R-\omega}$ [the difference of signs in $\Delta_L$ and $\Delta_R$ follows from the fact that Eq.~\eqref{hatcurtr} is formulated for the current flowing from the left to the right]; here the subscript ``TR'' refers to the transmission formalism. The transmission function of the nointeracting resonant level reads~\cite{nazarov2009}
\begin{align}
T(\omega)=\frac{\Gamma_L(\omega) \Gamma_R (\omega)}{(\omega-\epsilon_d)^2+\left[\Gamma_L(\omega)+ \Gamma_R (\omega) \right]^2/4},
\end{align}
where $\Gamma_\alpha (\omega)$ is the energy-dependent coupling strength to the lead $\alpha$. Here, as in the main text, we apply the boxcar-shaped coupling strength of the form
\begin{equation}
\Gamma_\alpha (\omega)=
\left\{\begin{array}{cc} \Gamma_\alpha & -W/2 \leq \omega \leq W/2 \\
0 & {\rm otherwise}
\end{array}\right..
\end{equation}

Within the correlation matrix formalism the heat current can be calculated as
\begin{align}
\dot{Q}_{\alpha,\text{CM}}=\frac{Q_\alpha (t+\delta t)-Q_\alpha (t)}{\delta t},
\end{align}
where $Q_\alpha (t)$ is the integrated heat calculated using Eq.~(22) from the main text and $\delta t$ is some small time interval (we choose ${\delta t=0.05}$); here the subscript ``CM'' refers to the correlation matrix formalism.

%
\begin{figure}
	\centering
	\includegraphics[width=0.9\linewidth]{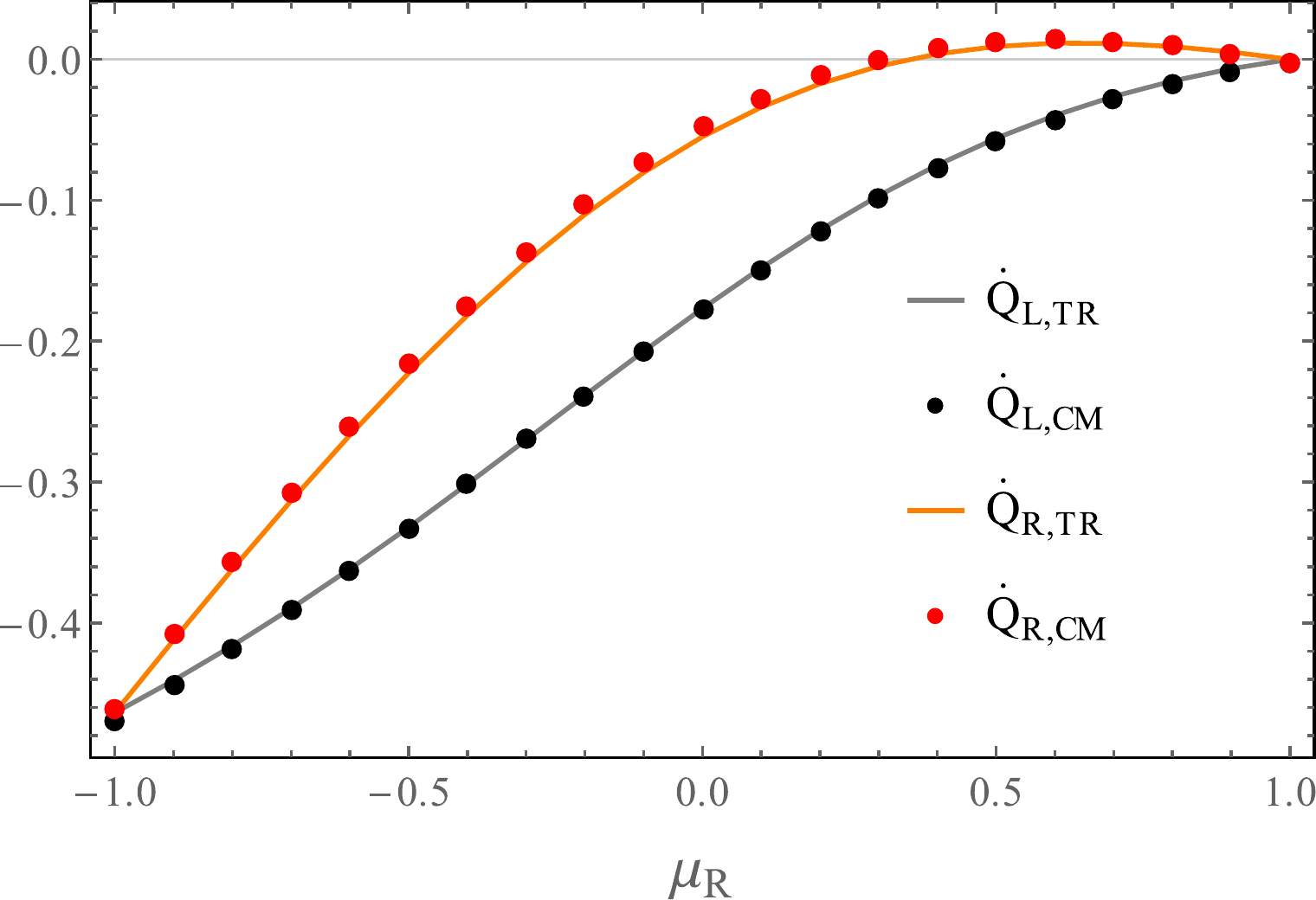}		
	\caption{Heat current calculated within the transmission matrix formalism (solid lines) and the correlation matrix formalism (dots) as a function of $\mu_R$ for ${n_d(0)=0}$, ${t=5}$, ${\delta t=0.05}$, ${\Gamma_L=\Gamma_R=1/2}$, ${\mu_L=1}$, ${\beta_L=\beta_R=3}$, ${\epsilon_{\alpha k}=(k-1)\Delta \epsilon-W/2}$, ${\Delta \epsilon=W/(K-1)}$, ${W=20}$ and ${K=256}$.}
	\label{fig:compst}
\end{figure}
%

Comparison of the results is presented in Fig.~\ref{fig:compst}. As one can clearly observe, both formalism agree well which confirms the relevance of the results from the main text for the thermodynamic limit.

\subsection{Scale-dependence of the transient dynamics} \label{sec:sc}
%
\begin{figure}
	\centering
	\includegraphics[width=0.88\linewidth]{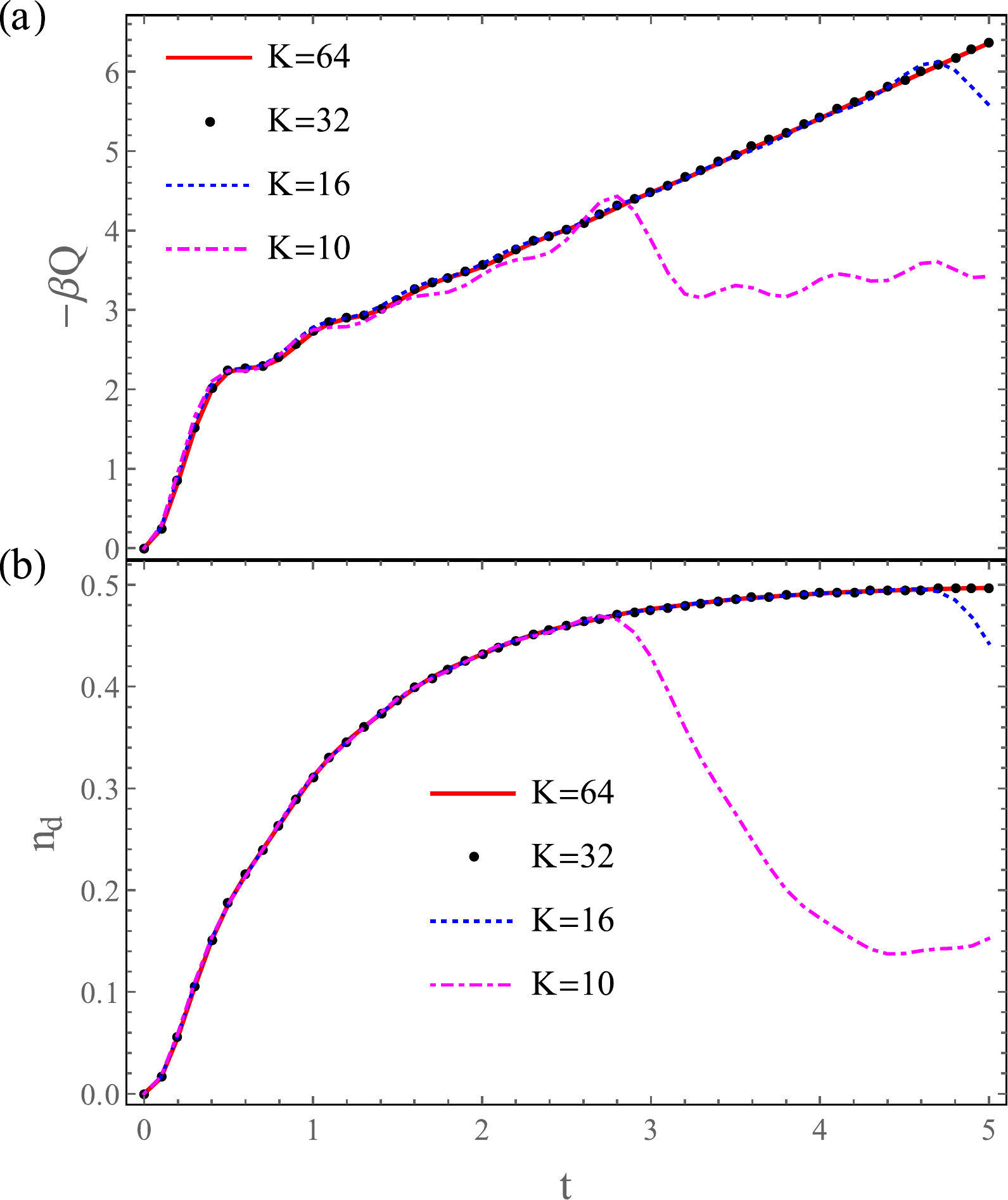}		
	\caption{The heat-related contribution to the entropy production baths ${-\beta Q=-\beta (Q_L+Q_R)}$ (a) and the occupancy of the central level $n_d$ (b) as a function of time for different number of sites $K$, ${n_d(0)=0}$, ${\Gamma_L=\Gamma_R=1/2}$, ${\mu_L=-\mu_R=1}$, ${\beta_L=\beta_R=\beta=3}$, ${\epsilon_{\alpha k}=(k-1)\Delta \epsilon-W/2}$, ${\Delta \epsilon=W/(K-1)}$ and ${W=20}$.}
	\label{fig:scalt5}
\end{figure}
%

%
\begin{figure}
	\centering
	\includegraphics[width=0.88\linewidth]{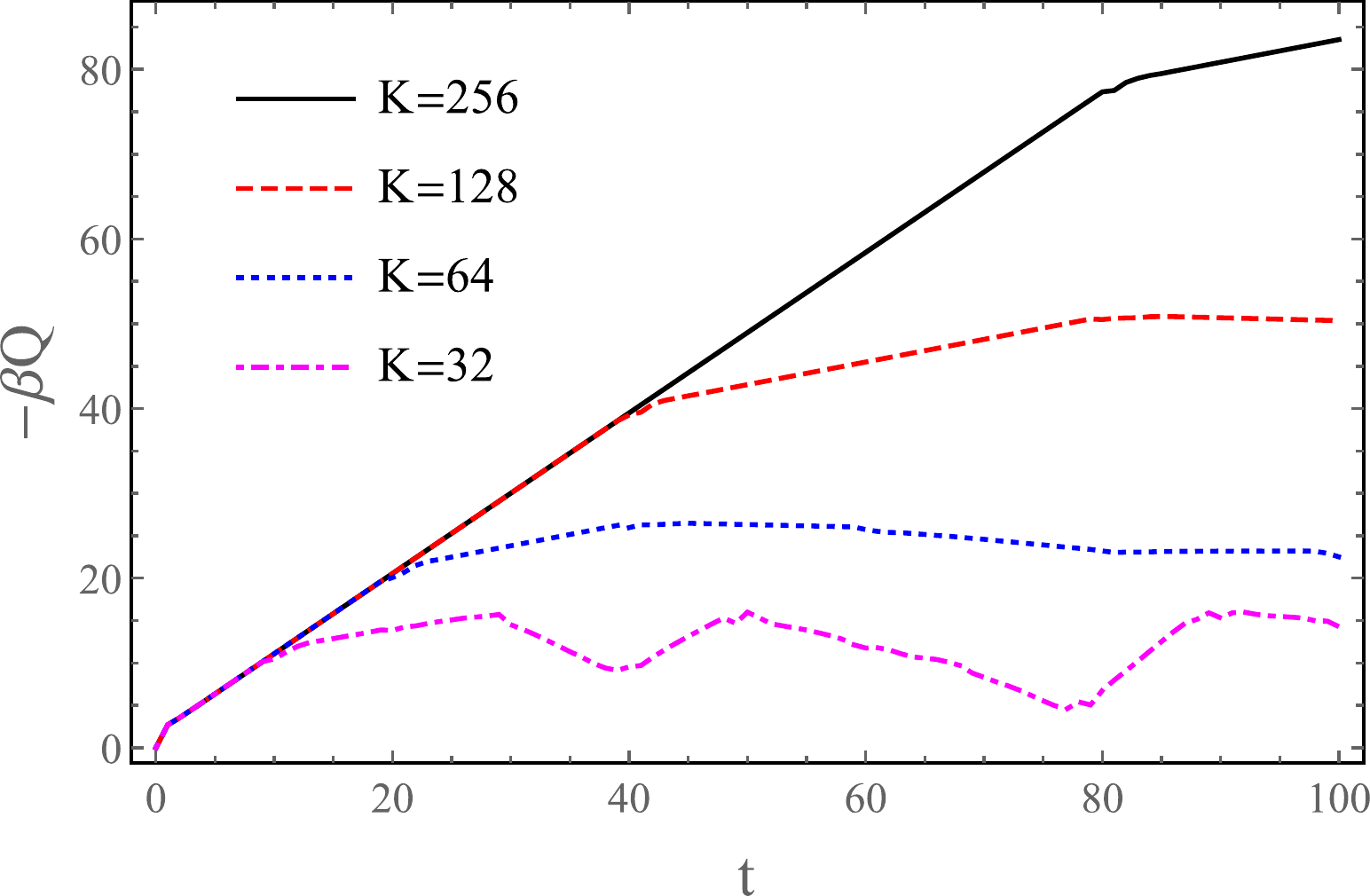}		
	\caption{The heat-related contribution to the entropy production baths ${-\beta Q=-\beta (Q_L+Q_R)}$ as a function of time for different number of sites $K$ and other parameters as in Fig.~\ref{fig:scalt5}.}
	\label{fig:scalt100}
\end{figure}
%

Here we analyze how the finite size of the baths affects the transient dynamics of the system. As Fig.~\ref{fig:scalt5} shows, for a short time the heat delivered to the baths follows the same track independently of the number of sites; the same is true for the occupancy of the central level. However, after a certain time, which depends on the number of sites, the heat saturates and exhibits irregular oscillations; this is a consequence of the finite size of the bath. Also the level occupancy diverges from the ``regular'' trajectory. This explains the sharp transition of the calculated thermodynamic quantities as a function of the number of sites in Fig.~2 in the main text: at ${t=5}$ the dynamics has already reached the saturation threshold for ${K \lessapprox 18}$, whereas for ${K \gtrapprox 18}$ the dynamics still follows the regular behavior corresponding to the steady state. As shown in Fig.~\ref{fig:scalt100}, for a sufficiently large number of sites in the bath the monotonous, linear increase of the heat is observed for times much longer than the duration of the initial transient state. This further demonstrates that the applied method, although confined to systems with finite baths, is well applicable to describe the steady state behavior of the system.

\subsection{Correlation between the baths}  \label{sec:infdec}
%
\begin{figure}
	\centering
	\includegraphics[width=0.9\linewidth]{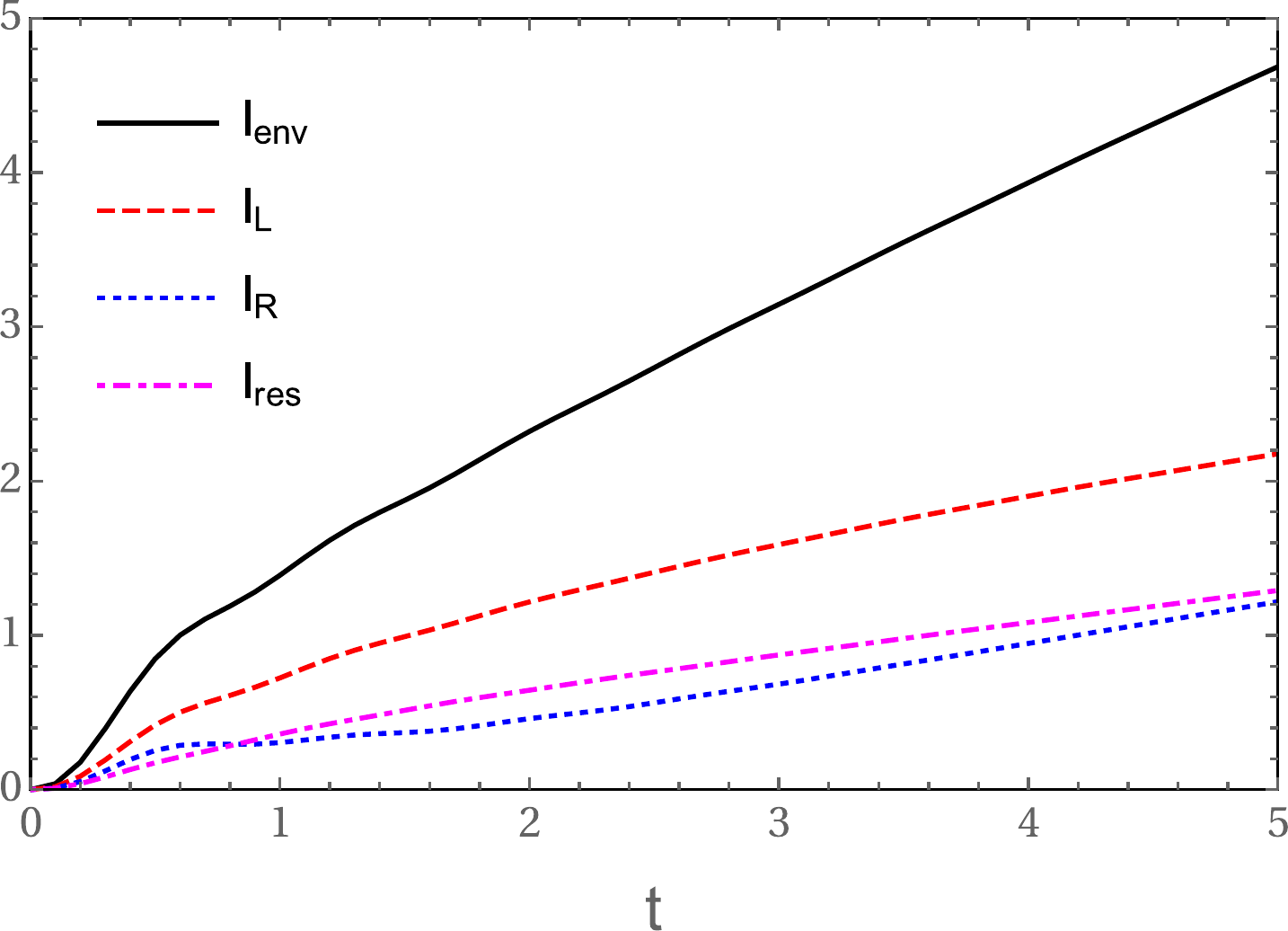}		
	\caption{Contributions to the mutual information $I_\text{env}$ as a function of time for ${K=256}$ and other parameters as in Fig.~\ref{fig:scalt5}.}
	\label{fig:contrinf}
\end{figure}
%
%
\begin{figure}
	\centering
	\includegraphics[width=0.9\linewidth]{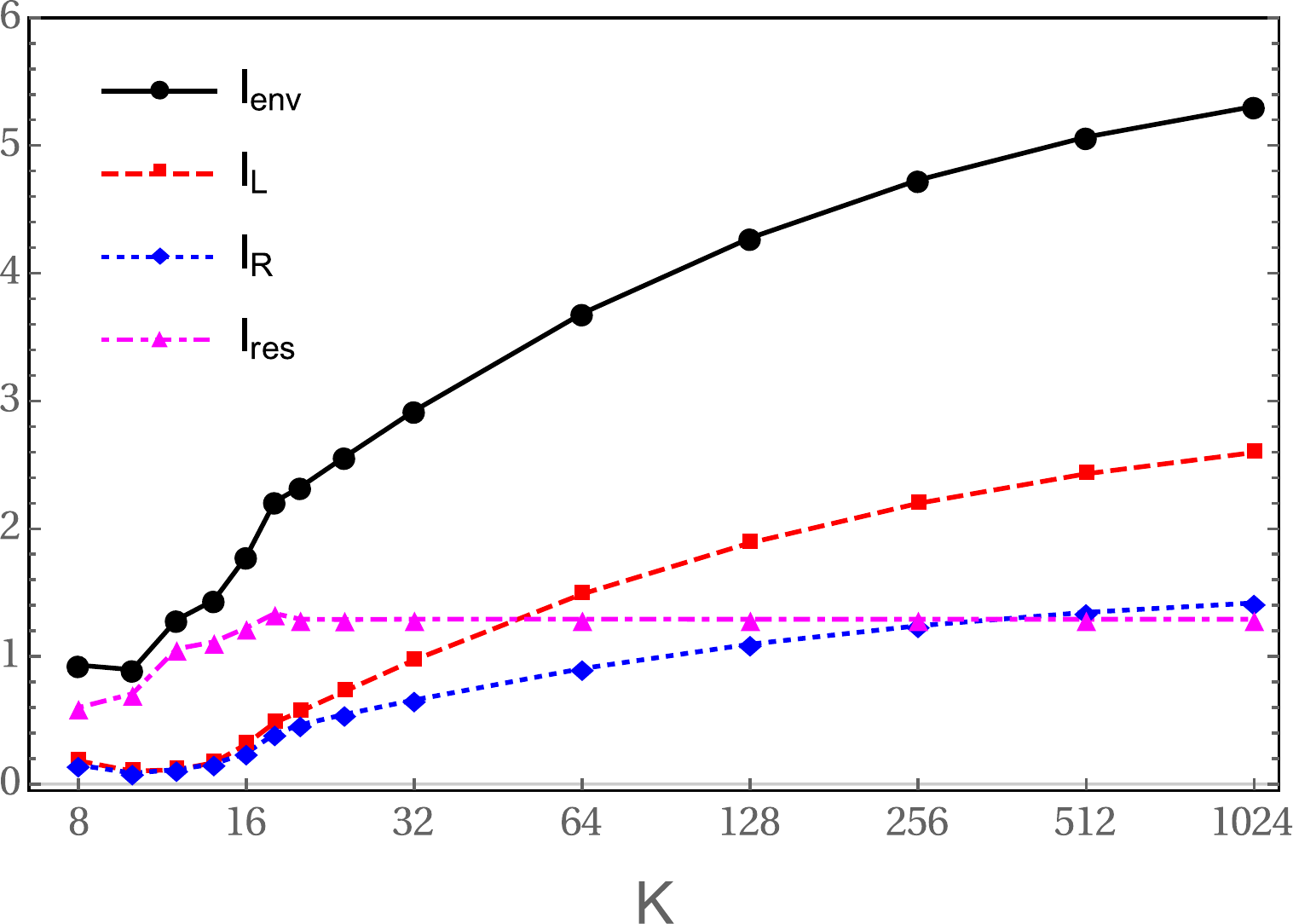}		
	\caption{Contributions to the mutual information $I_\text{env}$ as a function of the number of sites $K$ for ${t=5}$ and other parameters as in Fig.~\ref{fig:scalt5}. Results denoted by points, lines shown for eye guidance.}
	\label{fig:contrinfsc}
\end{figure}
%
Here we show that the mutual information of the intra-environment correlations is the result of both the correlation between the different baths and the intra-bath correlations. To demonstrate this, let us add and subtract $\sum_{\alpha} S_\alpha$ to Eq.~\eqref{infenv}, where ${S_\alpha =-\text{Tr} (\rho_\alpha \ln \rho_\alpha)}$ is the von Neumann entropy of the bath $\alpha$. As a result one obtains
\begin{align}
I_{\text{env}}=I_{\text{res}}+\sum_{\alpha} I_\alpha,
\end{align}
where 
\begin{align}
I_{\text{res}}=\sum_{\alpha} S_\alpha-S_E,
\end{align}
is the mutual information between the baths, whereas
\begin{align}
I_\alpha=\sum_{k} S_{\alpha k}-S_\alpha,
\end{align}
is the mutual information between degrees of freedom in the bath $\alpha$.

Figure~\ref{fig:contrinf} shows different contributions to the mutual information $I_{\text{env}}$ as a function of time for a given number of sites ${K=256}$. As one can see, the correlation between the baths and the intra-bath correlations are of similar order of magnitude and grow monotonously with time. Figure~\ref{fig:contrinfsc} demonstrates, on the other hand, the dependence of these contributions on the size of the bath for a fixed time ${t=5}$. As one can observe, the mutual information between the baths $I_{\text{res}}$ becomes scaling invariant for ${K \gtrapprox 18}$. In contrast, contributions due to the intra-bath correlations, as the mutual information $I_{\text{env}}$ itself, grow with the number of sites, which is due to decrease of the contribution $D_\text{env}$ (see Fig.~2 in the main text). We finally note that the correlations between the baths have been previously studied in Refs.~\cite{klich2009, song2011, song2012}, however, without connection to thermodynamics.